\documentclass[%
 reprint,
superscriptaddress,
 amsmath,amssymb,aps,
 prb,
]{revtex4-2}

\usepackage{graphicx}
\usepackage{dcolumn}
\usepackage{bm}
\usepackage{xcolor}
\usepackage{amsmath,amssymb}
\usepackage{float}
\usepackage{mathtools}

\begin{document}

\title{Topological edge and corner states in coupled wave lattices \\ in nonlinear polariton condensates}

\author{Tobias Schneider}
 \affiliation{Department of Physics and Center for Optoelectronics and Photonics Paderborn (CeOPP), Paderborn University, 33098 Paderborn, Germany}

\author{Wenlong Gao}
 \affiliation{EIT Institute for Advanced Study, Ningbo, China}

\author{Thomas Zentgraf}
\affiliation{Department of Physics and Center for Optoelectronics and Photonics Paderborn (CeOPP), Paderborn University, 33098 Paderborn, Germany}%
\affiliation{Institute for Photonic Quantum Systems (PhoQS), Paderborn University, 33098 Paderborn, Germany}

\author{Stefan Schumacher}
\affiliation{Department of Physics and Center for Optoelectronics and Photonics Paderborn (CeOPP), Paderborn University, 33098 Paderborn, Germany}%
\affiliation{Institute for Photonic Quantum Systems (PhoQS), Paderborn University, 33098 Paderborn, Germany}
\affiliation{Wyant College of Optical Sciences, University of Arizona, Tucson, AZ 85721, USA}%

\author{Xuekai Ma}
\affiliation{Department of Physics and Center for Optoelectronics and Photonics Paderborn (CeOPP), Paderborn University, 33098 Paderborn, Germany}%

\begin{abstract}
Topological states have been widely investigated in different types of systems and lattices. In the present work, we report on topological edge states in double-wave (DW) chains, which can be described by a generalized Aubry-Andr{\' e}-Harper (AAH) model. For the specific system of a driven-dissipative exciton polariton system we show that in such potential chains, different types of edge states can form. For resonant optical excitation, we further find that the optical nonlinearity leads to a multistability of different edge states. This includes topologically protected edge states evolved directly from individual linear eigenstates as well as additional edge states that originate from nonlinearity-induced localization of bulk states. Extending the system into two dimensions (2D) by stacking horizontal DW chains in the vertical direction, we also create 2D multi-wave lattices. In such 2D lattices multiple Su–Schrieffer–Heeger (SSH) chains appear along the vertical direction. The combination of DW chains in the horizontal and SSH chains in the vertical direction then results in the formation of higher-order topological insulator corner states. Multistable corner states emerge in the nonlinear regime.
\end{abstract}

\maketitle

\section{Introduction} 
Recent years have seen a surge on the study of phases of matter classified as topological in various physical platforms~\cite{xiao2010berry,hasan2010colloquium,qi2011topological,lu2014topological}. Despite being originally conceived in condensed matter systems for electrons, topological states have been reported in various photonic systems such as photonic crystals, coupled waveguides, metamaterials, optomechanics, and exciton-polaritons~\cite{ozawa2019topological}. Such topological states have shown prospect for bringing the coveted topologically protected robustness into various applications, for instance unidirectional signal propagation~\cite{haldane2008possible, wang2009observation}, optical communication, single mode lasing~\cite{st2017lasing}, vertical-cavity surface-emitting lasers (VCSELs)~\cite{yang2022topological}, quantum information processing~\cite{PhysRevLett.94.166802,alicea2011non}, just to name a few. Morover, photonic techniques were used to further extend the realm of topological states, some of which are not readily accessible in condensed matter systems, for instance Floquet topological insulators \cite{Rechtsman2013}, synthetic space topological insulators \cite{zilberberg2018photonic,lohse2018exploring}, Thouless pumping \cite{Citro2023}, and nonlinear topological insulators \cite{maczewsky2020nonlinearity}.     

Besides the study of novel types of topological structures, one aspect that deserves further attention is the  interplay between topology and nonlinearity. One platform of significant current interest in which both topology and nonlinearity can be conveniently investigated and controlled are so-called exciton polaritons. Exciton polaritons are partial-light partial-matter quasi-particles, originating from the strong coupling of excitons and photons in (planar) microresonators~\cite{kavokin2017microcavities}. The hybrid nature gives rise to a strong polariton-polariton interaction, which is one of the main factors for the realization of polariton condensation~\cite{deng2002condensation,kasprzak2006bose}, nonlinear functional polaritonic elements \cite{Luo2023,Zasedatelev2019,ma2020realization, luk2021all}, and enables the nonlinear control and trapping of macroscopically coherent polariton states in optically induced potential landscapes~\cite{Wertz2010,PhysRevB.88.041308,schmutzler2015,schneider2016exciton}. In tailored lattice structures, different topological edge states can be realized in the nonequilibrium polariton system such as in honeycomb lattices~\cite{klembt2018exciton,milicevic2015edge,Ma:20}, Lieb lattices~\cite{PhysRevLett.120.097401,PhysRevB.97.081103}, kagome lattices~\cite{gulevich2017exploring,PhysRevB.94.115437}, and Su–Schrieffer–Heeger (SSH) chains~\cite{st2017lasing,su2021optical,harder2021coherent}. It was also reported that the strong polariton nonlinearity can induce novel topological phenomena such as multistable vortices~\cite{ma2020realization,PhysRevLett.121.227404}, topologically protected solitons~\cite{Kartashov:16,gulevich2017exploring}, gap solitons~\cite{pernet2022gap}, coupling between corner modes in higher-order polariton insulators~\cite{PhysRevLett.124.063901}, bistable topological insulators~\cite{PhysRevLett.119.253904,zhang2019finite}, and the double-sided skin effect~\cite{PhysRevB.103.235306}.

\begin{figure*}[t]
  \centering
   \includegraphics[width=2.0\columnwidth]{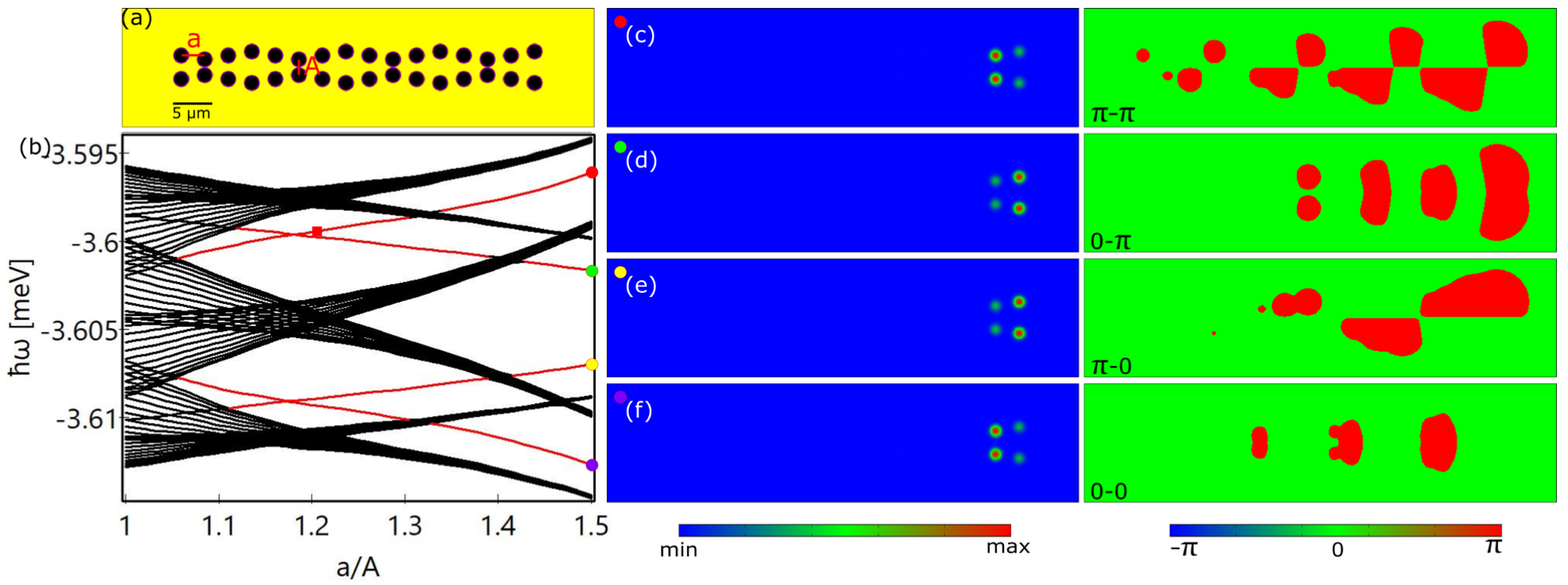}
  \caption{{\bf Edge states in double-wave (DW) chains.} (a) DW potential chain with $a=3$ $\mu$m, $A=2$ $\mu$m, and wave modulation amplitude $d=0.5$ $\mu$m. (b) Dependence of the energies (relative to the polariton dispersion minimum) of the linear eigenstates in the DW chain on the inter-wave separation $A$ with $a=3$ $\mu$m and $d=0.5$ $\mu$m fixed. It is calculated based on a longer DW chain that contains 52 potential wells in each wave, instead of the 16 shown in (a), to make the energy bands and band gaps easier to distinguish. Red lines indicate the topological edge states and black lines are the bulk states. (c-f) Amplitude (left) and phase (right) distributions of the edge states marked in (b).
  }
  \label{fig:1}
\end{figure*}

One prominent finding in quasi one-dimensional (1D) topological systems is the characteristic zero-dimensional (0D) localization of states on the edges. To systematically obtain such topological states in 1D systems, one possible route is by referring to the periodic table that classifies the ten discrete symmetry classes that coincide with the Altland-Zirnbauer (AZ) classification of random matrices~\cite{chiu2016classification}. For instance the topological group of the BDI class (one of the symmetry classes in the AZ table) is $\mathbb{Z}$, corresponding to a winding number in the first homotopy group. Such states, however, often find obstacles in their realization in photonic systems, due to their stringent symmetry requirements. A more feasible realization is the renowned SSH model, in which the introduction of both chiral and geometrical mirror symmetry endows a finer topological class beyond the AZ table. With its generalizations, the SSH model has been extensively studied in various photonic systems \cite{blanco2016topological,st2017lasing,zhao2018topological,parto2018edge,han2019lasing}. It has been reported that a two-dimensional (2D) SSH model supports higher-order topological insulator states~\cite{benalcazar2017quantized}, i.e., corner states, which recently triggered broad interest~\cite{PhysRevB.98.205147,PhysRevLett.122.233903,kim2020topological,wu2023higher}. A less intensely studied but not less promising route is the Aubry-Andr{\' e}-Harper (AAH) model~\cite{aubry1980analyticity} which translates into a Chern class in  synthetic space~\cite{ganeshan2013topological}. Such kind of model has recently been proposed in polariton systems in 1D by using optically induced potentials~\cite{PhysRevB.106.L220305}.  

In the present work, we introduce and investigate a specific class of lattices to realize topological edge states: double-wave (DW) chains as sketched in Fig.~1(a), which can be described in the framework of a coupled AAH model, i.e., cosinusoidal modulated potentials. First we discuss the topology of these structures based on a general tight-binding model Hamiltonian. Then we proceed to a specific physical realization in a planar semiconductor microcavity system with polaritonic excitations. We find that the edge states in the proposed DW chains become clearly distinguishable from the bulk states when the intra-wave separation between the potential wells is larger than the separation of the two waves, i.e., the inter-wave separation. By changing the specific structure of the waves, different edge states can be obtained, and they can even appear at different edges. Besides these edge states intrinsic to the structure, we find that the nonlinearity can transform specific bulk states into states that are localized at the edges. Moreover, in the nonlinear regime, the eigenstates on the edge with different phase distributions can be excited simultaneously, resulting in a final state where asymmetric edge states are stabilized by the nonlinearity. Several of these edge states can be stabilized for the same optical pumping parameters, resulting in a multistability that offers potential applications, e.g., in all-optical switching. Regularly stacking multiple identical DW chains into a 2D structure results in multiple SSH chains in the direction perpendicular to the DW chains. There, we demonstrate that the topology of the nontrivial DW chain and the nontrivial SSH chains leads to formation of corner states inside a band gap. Similar to the 1D case, different corner states can be stabilized by the same pump due to the nonlinearity. These results are obtained in a specific physical platform, the microcavity polariton system. However, it is important to emphasize that the topological structure and the topological states (edge states in 1D and corner states in 2D), introduced in the present work are of very general nature and can be realized in different physical implementations.

\begin{figure*}[htp]
  \centering
  \includegraphics[width=2\columnwidth]{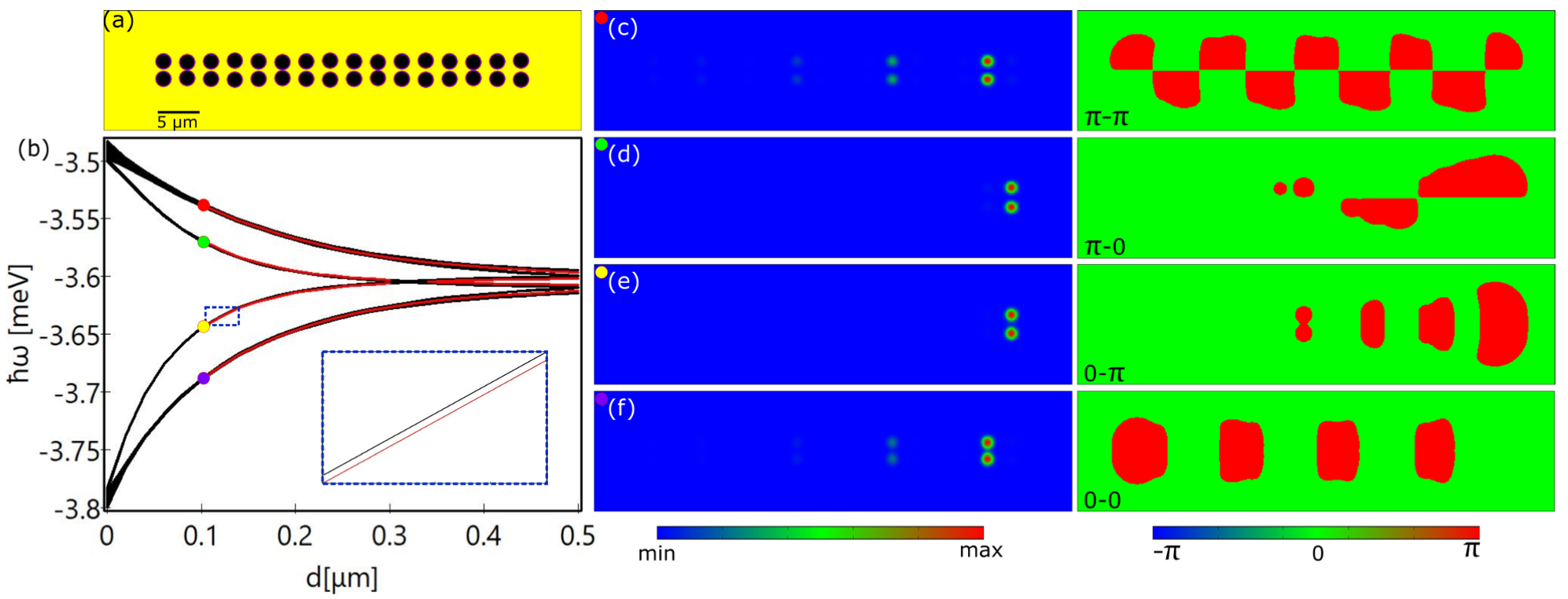}
  \caption{{\bf Influence of the modulation amplitude of the DW chain on the edge states.} (a) Distribution of the DW potential chain with $a=3$ $\mu$m, $A=2$ $\mu$m, and $d=0.1$ $\mu$m. (b) Dependence of the eigenenergies on the modulation amplitude $d$. Here $A=2$ $\mu$m and $a=3$ $\mu$m. Red lines indicate the topological edge states and black lines are the bulk states. (c-f) Density (left) and phase (right) distributions of the eigenstates in the DW chain presented in (a), corresponding to the symbols in (b), respectively.
  }
  \label{fig:s1}
\end{figure*}

\section{Lattices, Chern number, and theoretical model} 
Let us begin our discussion with the definition of the proposed DW structure and an analysis based on a generally applicable tight-binding (TB) model Hamiltonian. Further below we will then proceed to analyze a specific physical realization in the form of a planar semiconductor microcavity system and optical excitations therein. 

We assume 2D DW chains with potential wells that are distributed in the $x-y$ plane at $(x_n, {\pm}y_n)$ with $n\in[1,N]$. The positions of the potential wells are given by 
\begin{equation}
\begin{aligned}
\label{pillars}
    x_n=a(n-1), \ \ y_n=\frac{A}{2}+d-d\sin\left((n-1)\frac{\pi}{2}+\theta\right)\,.
\end{aligned}
\end{equation}
Here, $a$ is the separation of the neighbouring wells along the horizontal ($x$) direction, i.e., intra-wave separation, $A$ represents the separation of the closest wells in the upper and lower waves, i.e., inter-wave separation, and $d$ is the amplitude of each modulated wave. Figure \ref{fig:1}(a) shows the potential distribution with the wave constant $4a$, i.e., each four potential wells form a wave unit, and there are four periods along $x$ direction with a total of 16 potential wells. The separation of the upper and lower waves, center to center, is $A+2d$. The phase shift $\theta$ ($\in[0, 2\pi]$) in Eq.~\eqref{pillars} indicates the phase of the first (leftmost) well in the modulation and $\theta=0$ in Fig. \ref{fig:1}(a). In the DW chain, the profile of each individual potential well satisfies
\begin{equation}
\begin{aligned}
\label{potential}
    V(x,y)={\sum_n}V_0 e^{-\left(\frac{(x-x_n)^2+(y\pm y_n)^2}{w^2}\right)^{10}}.
\end{aligned}
\end{equation}
Here, $V_0$ is the potential depth and $w$ radius of each potential well.

Heuristically, this DW chain can be described by a TB model with Hamiltonian
\begin{equation}
\label{chain TBM}
\begin{multlined}
    \hat{H}=\sum_{n}^{N-1} (J_{n,n+1}A_n^{\dag}A_{n+1}+J_{n,n+1}B_n^{\dag}B_{n+1}+h.c.)\\
    +\sum_{n}^{N} (t_{n,n} A_n^{\dag}B_{n}+h.c.)
\end{multlined}
\end{equation}
Here only the fundamental mode of each well and their nearest neighbour couplings are considered. $X_n^{\dag} (X_n)$ is the creation (annihilation) operator on the $n$-th site ($X=A,B$) and $A (B)$ corresponds to the upper (lower) chain. Furthermore, the Hamiltonian $\hat{H}$ has mirror exchange symmetry with the corresponding exchange operator $\hat{M}$ defined as $\hat{M}\psi=\psi'=\begin{pmatrix} B_n\\ A_n\end{pmatrix}$. The Hamiltonian can be block-diagonalized onto the basis of $\hat{M}$ under the transformation $\mathbf{U}\hat{H}\mathbf{U}^{-1}=\hat{H}_{BD}$, in which $\hat{H}_{BD}=\begin{pmatrix}h_+ & 0 \\ 0 & h_-\end{pmatrix}$. 
Finally, if we assume the inter-wave couplings to take the form $t_{n,n}=t+\lambda \cos(2\pi \beta n+\theta)$ where $\beta =p/q$ is a rational parameter ($p$ and $q$ are co-prime), then $h_{+(-)}$ is inscribed as an AAH model, while $\hat{H}_{BD}$ governs two uncoupled AAH models with oppositely signed modulation. The Chern index associated with each energy band can be defined in the synthetic 2D parameter space $(\theta, k_x)$ for each mirror subspace as $C_{\pm,j} \mbox{ for } j=1,...,N$. Here $+(-)$ refers to the even(odd) subspace and $j$ is the index of bands whose energies are arranged in ascending order. Owing to the reversely signed modulation between the two subspaces, the relation $C_{+,j}=C_{-,N-j+1}$ is satisfied [see the Supplementary Materials (SM)]. Considering an individual wave unit with $N=4$, we have $C_{+,j}=C_{-,N-j+1}=\{1,-3,1,1 \}$ (the expression for the calculation of the Chern index is given in the SM), evidencing that the band gaps in such DW chains are topologically nontrivial.

\begin{figure*}[htp]
  \centering
   \includegraphics[width=2\columnwidth]{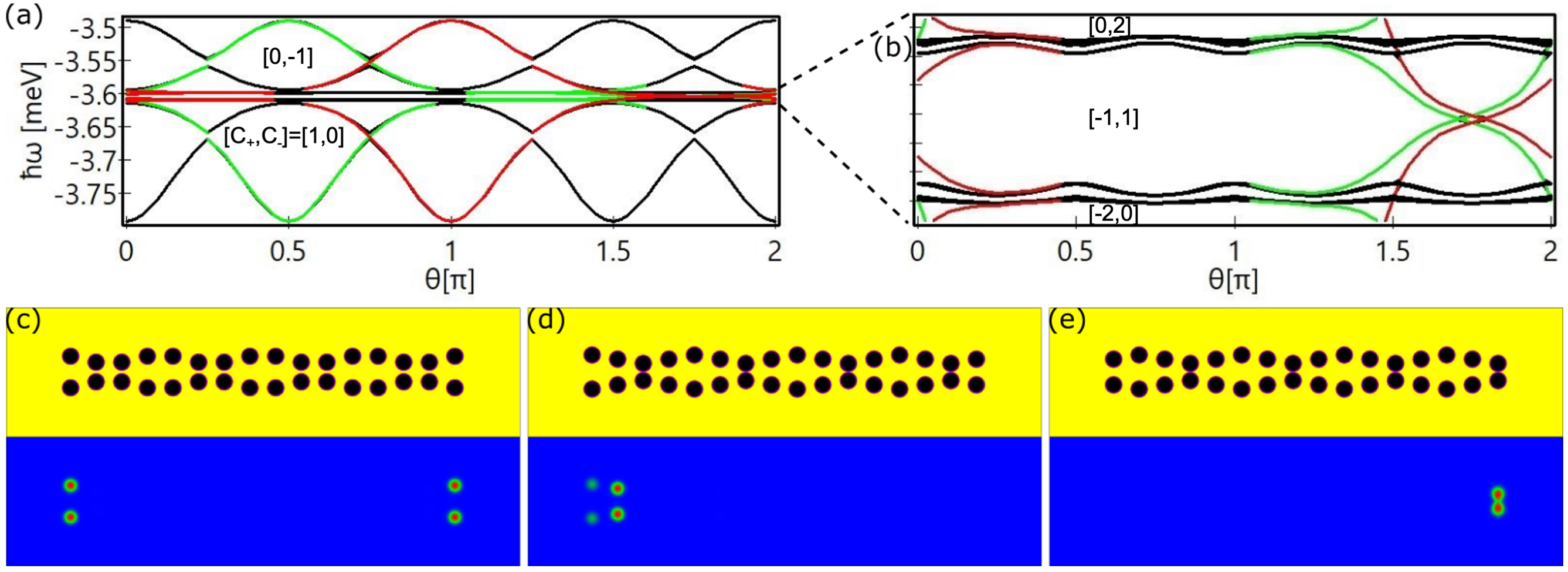}
  \caption{{\bf Edge states in differently structured DW chains.} (a) Dependence of the energies of the linear eigenstates on the phase shift $\theta$ of the chains, with $a=3$ $\mu$m, $A=2$ $\mu$m, and $d=0.5$ $\mu$m. Red lines indicate the right edge states, green lines indicate the left edge states, and black lines are the bulk states. (b) Enlarged view of the selected area in (a). The Chern numbers have been added into the corresponding band gaps. The detailed calculation can be found in the SM. (c-e) Different DW chain structures (upper panels) and amplitudes [same colorbar as in Fig.~1(c-f)] of the corresponding selected edge states (lower panels) at (c) $\theta=1.75\pi$, (d) $\theta=1.5\pi$, and (e) $\theta=\pi$.
  }
  \label{fig:2}
\end{figure*}

The TB model discussed above gives general and valuable insight into the topology of states expected in the proposed DW chains, similarly applicable to any physical realization. In the following, however, we will go a step further and analyze a specific physical realization of the DW structure and resulting topological states in a semiconductor microcavity polariton system. This analysis includes the dynamical system evolution to optically excite and populate the topological states as well as the interplay of topology and nonlinearity. 

To study the dynamical evolution of polariton condensates in the topologically nontrivial DW chains with resonant excitation, we consider a generalized Gross-Pitaevskii (GP) equation with loss and gain, i.e.,
\begin{equation}
\begin{aligned}
\label{GP_psi}
    i\hbar\frac{\partial \psi(\textbf{r},t)}{\partial t} =&\left[-\frac{\hbar^2}{2m}\nabla^2-i\hbar\frac{\gamma}{2}+g|\psi(\textbf{r},t)|^2+V(\textbf{r})\right]\psi(\textbf{r},t) \\
    \\
    &+E(\textbf{r},t)\,.
\end{aligned}
\end{equation}
Here, $\psi(\textbf{r},t)$ is the polariton wavefunction, $m=10^{-4}m_\text{e}$ (with free electron mass $m_\text{e}$) is the effective polariton mass, $\gamma$ is the loss rate in quasi-mode approximation \cite{Carcamo:20}, and $g$ is the nonlinearity coefficient representing the strength of the polariton interaction. $E(\textbf{r},t)$ is a coherent pump for the excitation of polaritons. $V(\textbf{r})$ represents the DW potential chain which can be fabricated in semiconductor microcavities in different ways \cite{schneider2016exciton}. For this physical realization, we chose $V_0=-5$ meV for the potential depth and $w=1$ $\mu$m for the radius of potential wells. These parameters are based on typical experiments in GaAs-based semiconductor microcavities~\cite{schneider2016exciton}. However, we note that the analysis and the GP model are similarly applicable to atomic condensates and nonlinear optics.

\section{Edge states in DW chains}
We start by studying the linear eigenstates of the DW chain. The eigenvalue problem can be obtained by substituting the ansatz $\psi(\textbf{r},t)=\psi(\textbf{r})e^{-i{\omega}t}$ into Eq.~\eqref{GP_psi} and neglecting the loss, gain, and nonlinearity. From the eigenenergies in Fig.~\ref{fig:1}(b), one can see that the gapped edge states (red lines) become more isolated from the bulk states (black lines) when the intra-wave separation $a$ is larger than the inter-wave separation $A$, akin to the appearance of the edge states in SSH chains when the intra-cell coupling is larger than the inter-cell coupling~\cite{RevModPhys.60.781}. When the edge states blend into the bulk states (energy bands), they become topologically trivial. Note that in order to clearly distinguish the energy bands and band gaps, the spectra shown in Fig.~\ref{fig:1}(b) are calculated for 13 unit cells (52 potential wells in each wave) instead of the case shown in Fig.~\ref{fig:1}(a) with only 4 unit cells. However, the topological properties in both cases remain the same [cf. Fig. S3(a) which is calculated with only 4 unit cells]. Since there are 8 potential wells in each unit cell, there must be 8 energy bands instead of 6 as shown in Fig.~\ref{fig:1}(b). Note that two energy bands are outside the shown energy range (one with a lower energy and the other with a higher energy, cf. Fig. S8).

In this configuration, all the edge states are located at the right edge of the chain. For each edge state four potential wells are occupied, and there are in total four different types of edge states as shown in Fig.~\ref{fig:1}(c-f). These edge states can be grouped into two subgroups, i.e., 0-states [Fig.~\ref{fig:1}(d,f)] and $\pi$-states [Fig.~\ref{fig:1}(c,e)]. For 0-states the phases in the equivalent wells in the two waves are the same, while for $\pi$-states there is a $\pi$-phase difference between them. In each group, there are two edge states where one shows the same phase for the last two wells in the same wave [Fig.~\ref{fig:1}(e,f)] and the other shows a $\pi$-phase jump [Fig.~\ref{fig:1}(c,d)]. Therefore, we denote the four edge states as 0-0 state [Fig.~\ref{fig:1}(f)], $\pi$-0 state [Fig.~\ref{fig:1}(e)], 0-$\pi$ state [Fig.~\ref{fig:1}(d)], and $\pi$-$\pi$ state [Fig.~\ref{fig:1}(c)]. It is worth noting that in the present work both types of localized states [Fig.~\ref{fig:1}(c-f)] are defined as edge states. That is, the edge states generally refer to the localized states that emerge in only the edge unit cells with the bulk unit cells empty. Besides the separation of the two waves, the properties of the edge states are also related to the modulation amplitude of the waves $d$. As $d$ decreases, the inter-wave interaction in the whole chain becomes stronger, leading to edge states that are less distinguishable from the bulk states as shown in Fig.~\ref{fig:s1}(b). Here, the modulated waves cannot be easily recognized, that is, the two chains are almost parallel to each other [Fig. \ref{fig:s1}(a)]. As a result, the 0-0 state and $\pi$-$\pi$ state slightly permeate through the bulk potential wells [Fig. \ref{fig:s1}(c,f)] and move deep into the bulk states [Fig. \ref{fig:s1}(b)], whereas the $\pi$-0 and 0-$\pi$ edge states become more localized in the edge potential wells [Fig. \ref{fig:s1}(d,e)]. These two states remain in the band gaps as can be seen from the inset in Fig. \ref{fig:s1}(b), although the broad frequency range of Fig. \ref{fig:s1}(b) makes them difficult to be distinguished from the bulk states. Such edge states remain as more potential periods are added to the left or right side of the chain. We note that our study shows representative results for a specific choice of system parameters. Band gaps can be broadened significantly by increasing the potential depth and hopping between adjacent potential wells can be enhanced by moving them closer together~\cite{schneider2016exciton}.

The reason why the edge states in Figs.~\ref{fig:1} and \ref{fig:s1} appear at the right edge rather than the left one is the specific arrangement of the potential wells. How the potential wells are arranged determines which edge (left or right or both) is occupied as illustrated in Fig.~\ref{fig:2} in which the phase shift $\theta$ is systematically varied in panels (a) and (b) from zero to 2$\pi$. When $\theta=1.75\pi$, the chain becomes symmetric along $x$ direction [Fig.~\ref{fig:2}(c)], and consequently both edges are occupied. In this case there appear two pairs of degenerate edge states [cf. Fig.~\ref{fig:2}(b)], because the wavefunctions at the two edges are spatially well separated from each other with virtually zero overlap, resulting in the degeneracy of the 0-0 and 0-$\pi$ states and $\pi$-0 and $\pi$-$\pi$ states. When $\theta=1.5\pi$ [Fig.~\ref{fig:2}(d)], the chain is spatially reversed in $x$ direction, compared with the one in Fig.~\ref{fig:1}(a). Under this condition the four edge states relocate to the left edge of the chain. When the chain is in the configuration shown in Fig.~\ref{fig:2}(e) where $\theta=\pi$, only two edge states that are localized in the rightmost wells appear. From the results presented in Fig.~\ref{fig:2} (see also the extended version of this figure, Fig. S1, in the SM), it is clear that the edge states arise only when the edge wells are not at the exact antinodes of the DW structure, i.e., only when $\sin\left[(n-1)\frac{\pi}{2}+\theta\right]\neq0$ in Eq.~\eqref{fig:3}. That is why the left edge in Fig.~\ref{fig:1} and Fig.~\ref{fig:2}(e) and the right edge in Fig.~\ref{fig:2}(d) are unoccupied. The appearance of the edge states shown in Fig.~\ref{fig:2} is consistent with the inscribed Chern number calculated from the tight-binding model, Eq.~\eqref{chain TBM}, and thus referred to the bulk-edge correspondence \cite{hatsugai2016bulk}. The edge states can also be reconstructed by directly adding or cutting specific potential wells at the left or right side of the chain in Fig.~\ref{fig:1}(a). Several examples are given in the SM (Fig. S2).

\section{Multistable edge states}
To optically excite the edge states and study their nonlinear dynamics, a resonant continuous-wave (CW) pump $E(\textbf{r},t)=E_0{e^{-i{\omega}t}}$, where $E_0$ is the constant amplitude and $\omega$ is the frequency, is used to continuously drive the system until the stationary solution is reached. In our calculations, the polaritons have a loss rate of $\gamma=0.005$ ps$^{-1}$ and a nonlinearity coefficient of $g=2$ $\mu$eV $\mu$m$^{2}$. The time evolution of Eq.~(1) is computed by using a 4th order Runge-Kutta method with zero boundary conditions. The stability of the stationary solutions is tested by adding white noise into the solutions and letting them evolve over $10\,\mathrm{ns}$.

\begin{figure}[t]
  \centering
   \includegraphics[width=1.0\columnwidth]{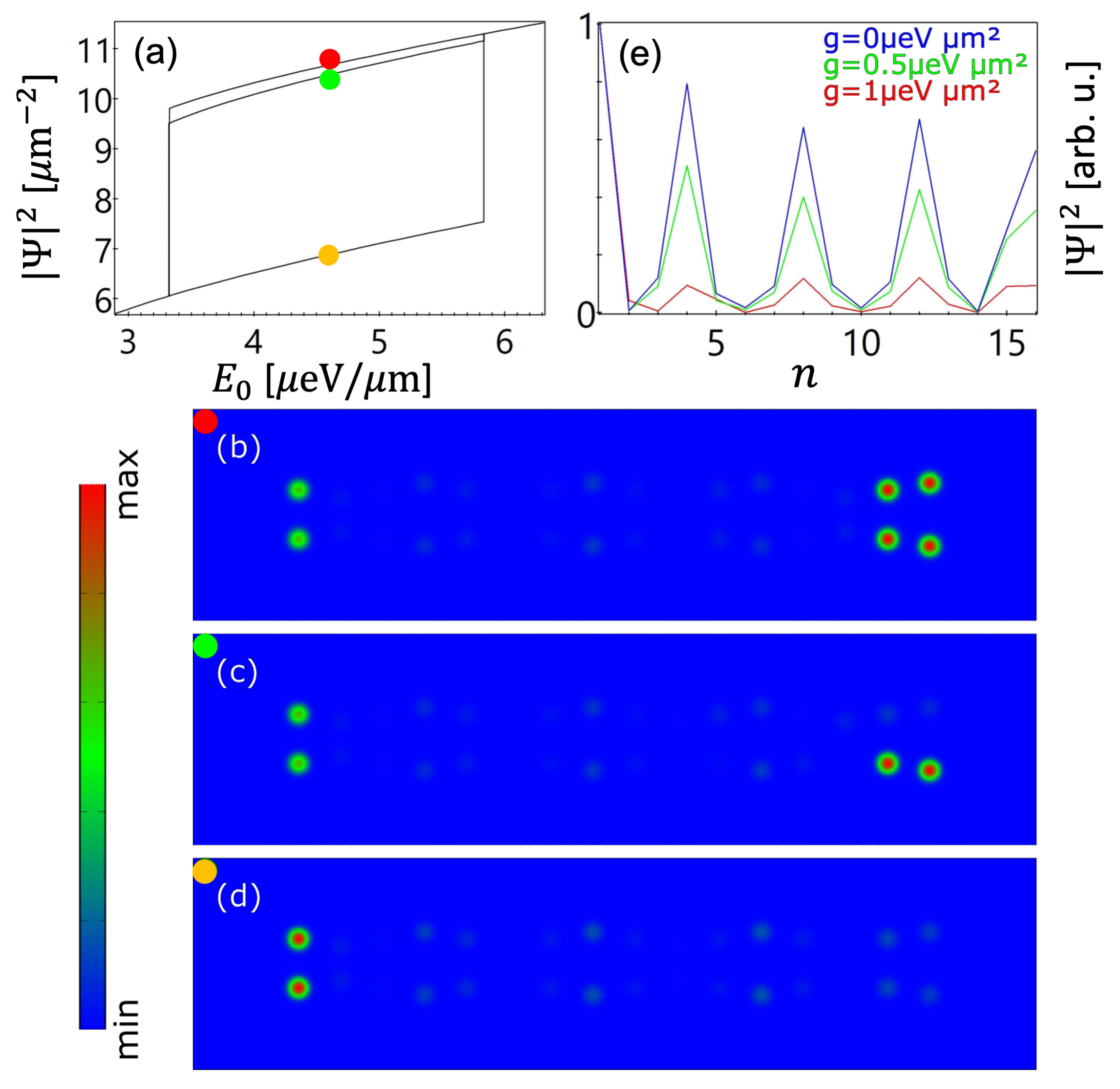}
  \caption{\textbf{Nonlinearity and multistable edge states.} (a) Dependence of the peak densities of the edge states on the amplitude of the continuous wave, spatially homogeneous pump with photon energy $\hbar\omega=-3.5995$ meV [indicated by the red square in Fig.~\ref{fig:1}(b) for $a=3$ $\mu$m, $A=2.5$ $\mu$m, $d=0.5$ $\mu$m, and $\theta=0$]. (b-d) Density profiles of multistable edge states corresponding to the points marked in panel (a). (e) Normalized peak densities in each potential well from the upper wave of the state in (d) for different nonlinear coefficients $g$.
  }
  \label{fig:3}
\end{figure}

\begin{figure*}[t]
  \centering
   \includegraphics[width=2\columnwidth]{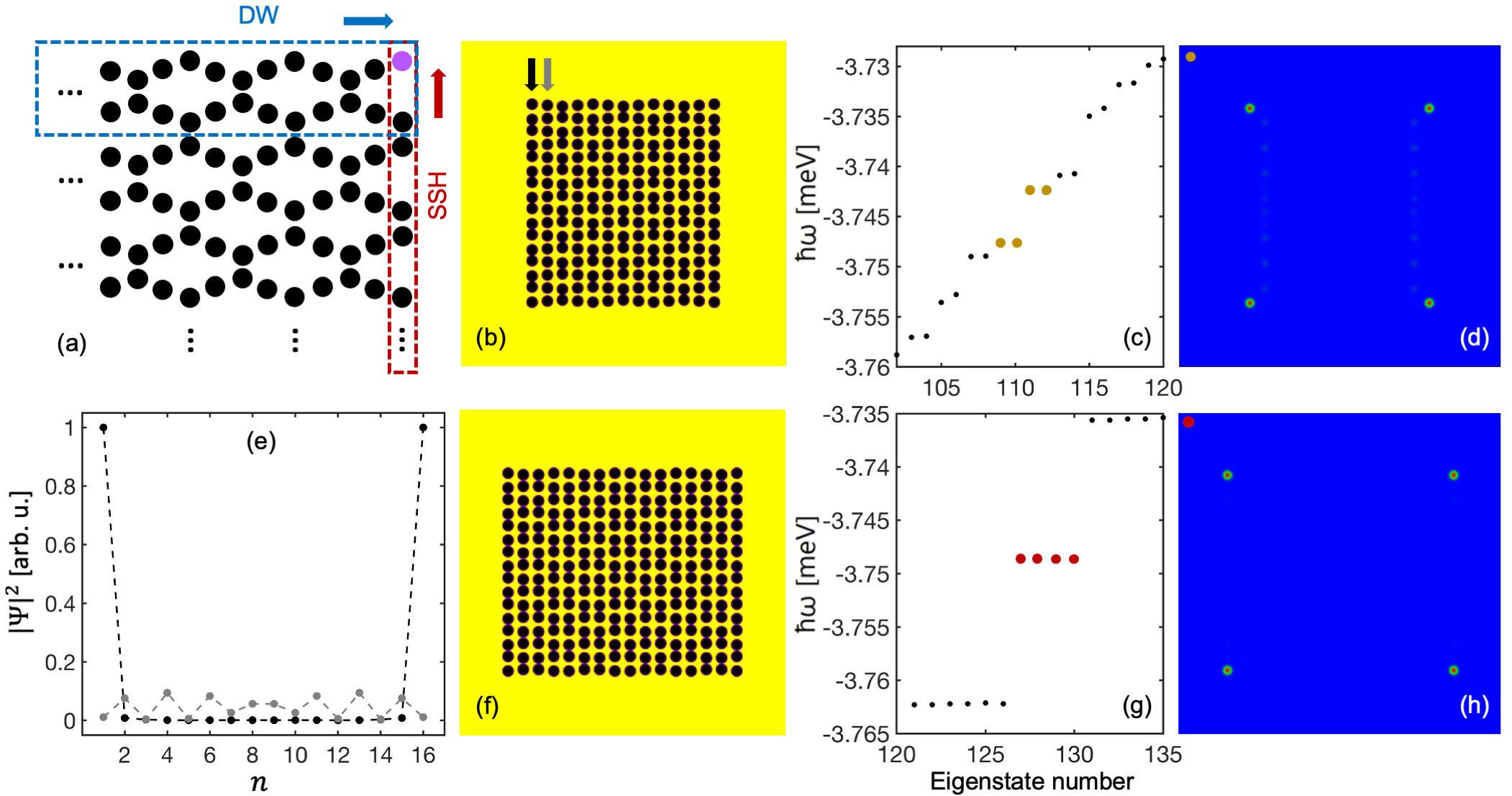}
  \caption{\textbf{Corner states.} (a) Schematic of multiple DW chains arranged along the vertical direction and the illustration of the formation of a corner state. (b) A 2D lattice pattern with $a=2.8$ $\mu$m, $A=2$ $\mu$m, and $d=0.2$ $\mu$m, and the unit DW chain has the similar structure of that in Fig. \ref{fig:s1}(a). (c) Selected linear eigenstates (energy vs. number) of the lattice in (b). (d) Density distributuion of the highlighted eigenstate 109 in (c). (e) Normalized peak densities (in each potential well) of the condensates, collected respectively from the two columns indicated by the two arrows in (b). (f) A 2D lattice pattern with $a=2.8$ $\mu$m, $A=2$ $\mu$m, and $d=0.2$ $\mu$m, and the unit DW chain has the same structure of that in Fig. \ref{fig:2}(c). (g) Selected linear eigenstates (energy vs. number) of the lattice in (f). (h) Density distributuion of the highlighted eigenstate 127 in (g).}
  \label{fig:4}
\end{figure*}

From Fig.~\ref{fig:1}(b), one can see that two edge states are degenerate at around $a/A=1.17$. Therefore, we choose the inter-wave separation of the chain at $a/A=1.2$ , where the 0-$\pi$ and $\pi$-$\pi$ states are very close to each other, and tune the photon energy of the pump to $\hbar\omega=-3.5995$ meV [see the red square in Fig.~\ref{fig:1}(b)]. In this scenario, because of the finite polariton lifetime and nonlinearity, the 0-$\pi$ and $\pi$-$\pi$ states can be simultaneously excited, and their phase difference leads to destructive interference of the condensate density in the upper wave [see the density at the right edge in Fig.~\ref{fig:3}(c)]. Meanwhile, the left edge is also occupied by the condensate; however, this state is not one of the linear eigenstates as known from Fig.~\ref{fig:1}. This edge state can even be separately excited as shown in Fig.~\ref{fig:3}(d) especially when the pump intensity is lower. To illustrate this point further, we use the left-edge state shown in Fig.~\ref{fig:3}(d) as the initial condition and gradually decrease the nonlinear coefficient while keeping the pump fixed. It is clearly seen from Fig.~\ref{fig:3}(e) that the nonlinearity strongly affects this edge state. With the decrease of the nonlinearity, the condensate starts to move to the bulk of the system, forming a progressively decreasing density distribution from left to right [the blue line in Fig.~\ref{fig:3}(e)], which is similar to the linear eigenstates with frequency below the 0-$\pi$ edge state [see the eigenstates in the SM Fig. S3]. In conclusion the nonlinearity pushes the bulk states to the edge, leading to the formation of nonlinearity-enhanced edge states, analogous to surface gap solitons~\cite{PhysRevLett.96.073901}. As the pump intensity increases [Fig.~\ref{fig:3}(a,b)], the right edge can be fully excited, which means only one of the intrinsic edge states is excited here. If we excite the edge state in the lattice with $a/A=1.5$, where the edge states are more isolated from the bulk states, by choosing the photon energy of the pump between the 0-$\pi$ and $\pi$-$\pi$ states, it is possible that the condensate is loaded only into the wells at the right edge, leaving the potential wells at the left edge empty [see Fig. S4 in the SM]. From Fig.~\ref{fig:3}(a), one can see that all these edge states can be stable in a broad pump intensity range. We note that this finding of edge-state multistability is of general nature in our structure as it can occur at different lattice constants and pump frequencies [cf. Fig. S4].

\begin{figure*}[t]
  \centering
   \includegraphics[width=2\columnwidth]{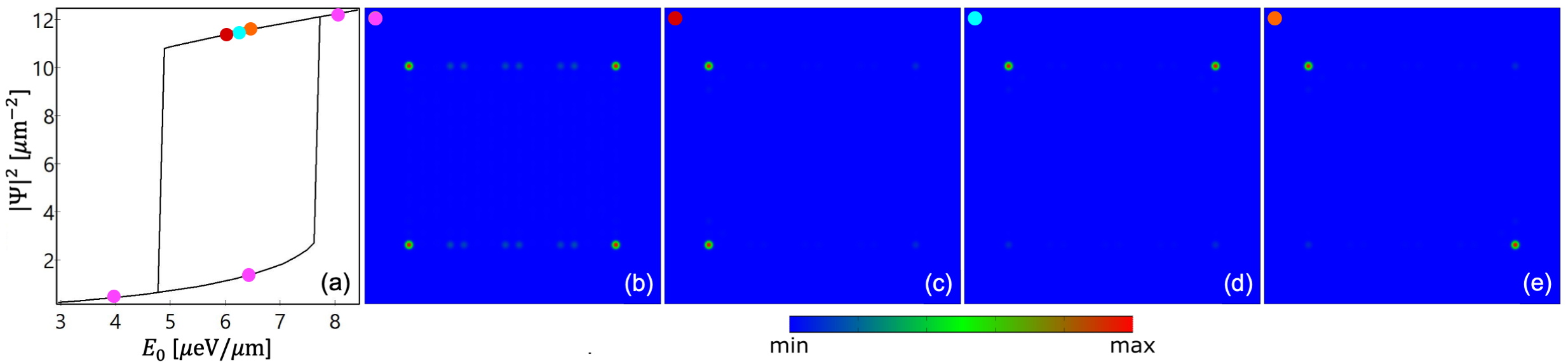}
  \caption{\textbf{Multistable corner states.} (a) Dependence of the peak densities of the corner states in the lattice presented in Fig.~\ref{fig:4}(f) on the amplitude of the continuous wave, spatially homogeneous pump with photon energy $\hbar\omega=-3.7449$ meV. (b-e) Density profiles of multistable corner states corresponding to the points marked in (a).
  \label{fig:6}}
\end{figure*}

\section{Corner states in multi-wave lattices}
Although one of the corners of the DW chain in Fig.~\ref{fig:3}(c) is occupied, it is not a higher-order topological insulator state, because it is induced by nonlinearity instead of a system's eigenstate. To realize higher-order topological insulators, we extend a chain along the vertical direction, i.e., perpendicular to the chain direction, with multiple repetitions to generate a 2D lattice. Examples can be found in Fig.~\ref{fig:4}(a,b,f), in which the separation of the DW chains is always the same as the wave separation in each individual DW chain, i.e., $A+2d$. From the schematic in Fig.~\ref{fig:4}(a), one can see that the DW chains along the horizontal direction allow the polaritons to condense in the entire right-edge potential wells as studied above. However, the potential distribution in every column of the lattice can be regarded as an SSH chain. When the SSH chain at the right (or left) edge has a larger inter-cell coupling as illustrated in Fig. \ref{fig:4}(a), another kind of topologically nontrivial edge state is expected. As a consequence, the combination of the DW chains in the horizontal direction and the SSH chains in the vertical direction can give rise to the emergence of a corner state, that is, the polariton condensate can be solely loaded into the corner potential well marked in purple in Fig.~\ref{fig:4}(a).

For the realization of such corner states, we first choose a DW chain with a smaller inter-wave separation to enable almost all the polaritons to condense in the rightmost (or leftmost) potential wells, i.e., reducing the density of the condensate in the potential wells next to the edges, referring to the edge states shown in Fig.~\ref{fig:s1}(d,e). The reason for reducing the condensate population in the bulk potential wells as much as possible is that in the SSH chain next to the right edge in Fig.~\ref{fig:4}(a), the inter-cell coupling is prominently reduced, so that the edge states in this potential column may vanish and the condensates in the bulk potential wells would hinder the formation of a clean corner state. By extending a DW chain with a similar structure of the one shown in Fig.~\ref{fig:s1}(a) (but symmetric distribution along the horizontal direction) into 2D (8 repetitions along the vertical direction), a lattice containing multiple potential waves is created as shown in Fig.~\ref{fig:4}(b). Figure~\ref{fig:4}(c) presents a part of the eigenstates including the corner states as highlighted. It can be seen that the band gap here is not very prominent due to the small wave modulation $d$ as illustrated in Fig.~\ref{fig:s1}(b). What is intriguing is that the four corner states share two different energies instead of one. This is because there are still polariton condensates being loaded into the two columns adjacent to the two vertical edges, which can be vaguely seen in Fig.~\ref{fig:4}(d). A clearer picture is presented in Fig. \ref{fig:4}(e) where the peak condensate densities in each potential well are extracted from the left-edge chain and the chain next to it, indicated by the two arrows in Fig.~\ref{fig:4}(b). A small amount of condensate appears in the bulk potential wells, and the energy difference of the four corner states originates from the phase difference ($0$ or $\pi$, see Fig. S5 in the SM) of the condensates in these two adjacent chains. 

To completely remove the influence of the weak polariton density in the bulk potential wells on the corner states, we focus on the DW distribution shown in Fig.~\ref{fig:3}(c) in which only two edges are occupied. Extending this chain along the vertical direction gives a 2D lattice pattern shown in Fig. \ref{fig:4}(f). In this case, because of the vanishing of the polaritons in the bulk potential wells, the condensates in the four corner potential wells becomes completely disconnected [Fig. \ref{fig:4}(h)], so that the four edge states are degenerate in energy [Fig. \ref{fig:4}(g)]. The band gap here is broader than that in Fig.~\ref{fig:4}(c) because of the increased wave modulation $d$. Apart from the 0D corner states, there exist also 1D edge states in such 2D lattices, examples can be found in Fig.~S6 in which the dependence of the topological states and the wave modulation $d$ is also illustrated. It is known that when the intra-cell coupling is stronger than the inter-cell coupling in an SSH chain, the topological edge states disappear. Therefore, if the top single-wave chain in Fig.~\ref{fig:4}(a) is removed, the corner state will be erased accordingly from the system. An example is given in the SM Fig. S5, in which the lattice is obtained by extending the DW chain in Fig.~\ref{fig:2}(e) into 2D. In this scenario, a SSH chain with a larger intra-cell coupling is formed at the right edge, hence the corresponding band gap is empty.

The corner state presented in Fig.~\ref{fig:4}(h) can also be optically excited by a homogeneous pump with a near resonant photon energy. Similar to the case studied in the 1D DW chain in Fig.~\ref{fig:3}, the nonlinearity can support more than one stable corner state as illustrated in Fig.~\ref{fig:6}. That is, besides the intrinsic individual eigenstate with four corners [Fig.~\ref{fig:6}(b)], the combination of any two different eigen corner states can also be excited due to the nonlinearity, resulting in new corner states with only two corners being excited as can be seen in Fig.~\ref{fig:6}(c-e). Due to the missing two corners, these states have a higher shared density than that of the four-corner state under the same pump in the multistability region [see Fig. \ref{fig:6}(a)]. Outside this region, only the four-corner states exist.

\section{Conclusion and discussion}
In summary, we report on different topological edge states in polaritonic DW chains. Their topological nature is evidenced by the nonzero Chern numbers in the linear regime. Under specific pumping conditions more than one edge state can be stabilized, including the combination of different intrinsic edge states as well as nonlinearity-enhanced ones. These states can appear at different edges for differently structured DW potentials. We also demonstrate that SSH chains can occur in multiple coupled DW chains (2D lattices), ascribed to the spatial wave distribution of the potential wells, and their direction is perpendicular to that of the DW chains, which support 0D localization in a band gap. Multistable corner states are also studied in the nonlinear regime. To demonstrate the robustness of the edge and corner states, we include a strong defect, by completely removing a corner potential well, in both 1D and 2D cases to study its influence on these interesting topological states and find that the edge and corner states are very robust and show stable time evolution in the presence of a defect, without evolving into bulk states (see Fig. S7 in the SM).

Although implemented in a specific system, our theoretical model and main results are of general nature. Similar observations of topologically protected edge and corner states in these  structures are expected in other physical realizations such as solid-state and photonic systems utilizing the advantages of our structure. For example, the 1D DW chains, comparing with the SSH and AAH chains, enable the emergence of different edge states and their combinations can be stabilized by nonlinearity, giving rise to multistable edge states. The switching between the multistable edge states for example, which is beyond the scope of this work, could benefit the development of all-optical switches. In polariton condensates, the topological edge states in the SSH chains occur at higher-order modes (dipoles) with higher energy, instead of fundamental modes~\cite{st2017lasing,su2021optical,harder2021coherent}, while in the DW chains the topological edge states appear in the fundamental modes with lower energy.

\begin{acknowledgments}
This work was supported by the Deutsche Forschungsgemeinschaft (DFG) (No. 467358803 and 519608013) and by the Paderborn Center for Parallel Computing, PC$^2$. S.S. and T.Z. further acknowledge financial support by the DFG
Collaborative Research Center TRR 142 (No. 231447078, projects B09 and A04).
\end{acknowledgments}

\end{document}


\maketitle

\newpage

\section*{I. Configuring different edge states}
\begin{figure}[H]
  \centering
   \includegraphics[width=1.0\columnwidth]{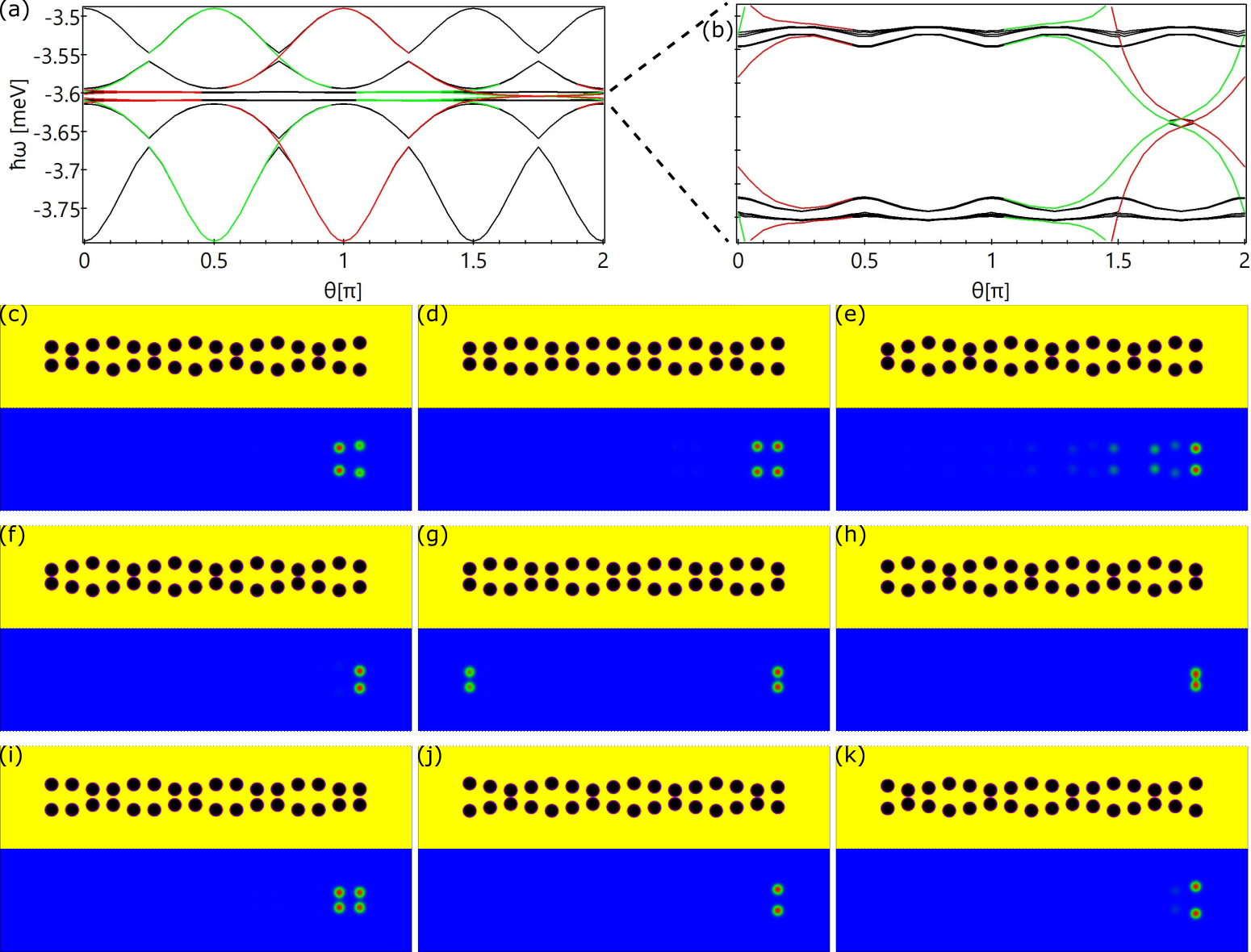}
  \caption{{\bf Edge states in differently structured double-wave chains.} (a) Dependence of the eigenenergies of the linear modes on the phase shift of the chains $\theta$. Here $a=3$ $\mu$m, $A=2$ $\mu$m, and $d=0.5$ $\mu$m. Red lines indicate the right edge states, green lines indicate the left edge states, and black lines are the bulk states. (b) Enlarged view of the selected area in (a). (c-k) Profiles of the chain (upper) and the corresponding density of the selected edge states (lower) at different phase shifts: (c) $\theta=0.1\pi$, (d) $\theta=0.25\pi$, (e) $\theta=0.45\pi$, (f) $\theta=0.5\pi$, (g) $\theta=0.75\pi$, (h) $\theta=\pi$, (i) $\theta=1.25\pi$, (j) $\theta=1.525\pi$, and (k) $\theta=1.9\pi$.
  }
  \label{fig:s2}
\end{figure}

Figure \ref{fig:s2} is the extended version of Fig. 2, where more examples of the chain distributions and the corresponding edge states are shown. It is clearly seen that the edge states appear as long as the edge potential wells are the most or lest separated ones in the double waves, even if the phase shift $\theta$ is an arbitary value [see Fig. \ref{fig:s2}(j)]. 

\newpage

\section*{II. Edge states in double-wave chains with different lengths}
\begin{figure}[H]
  \centering
   \includegraphics[width=1.0\columnwidth]{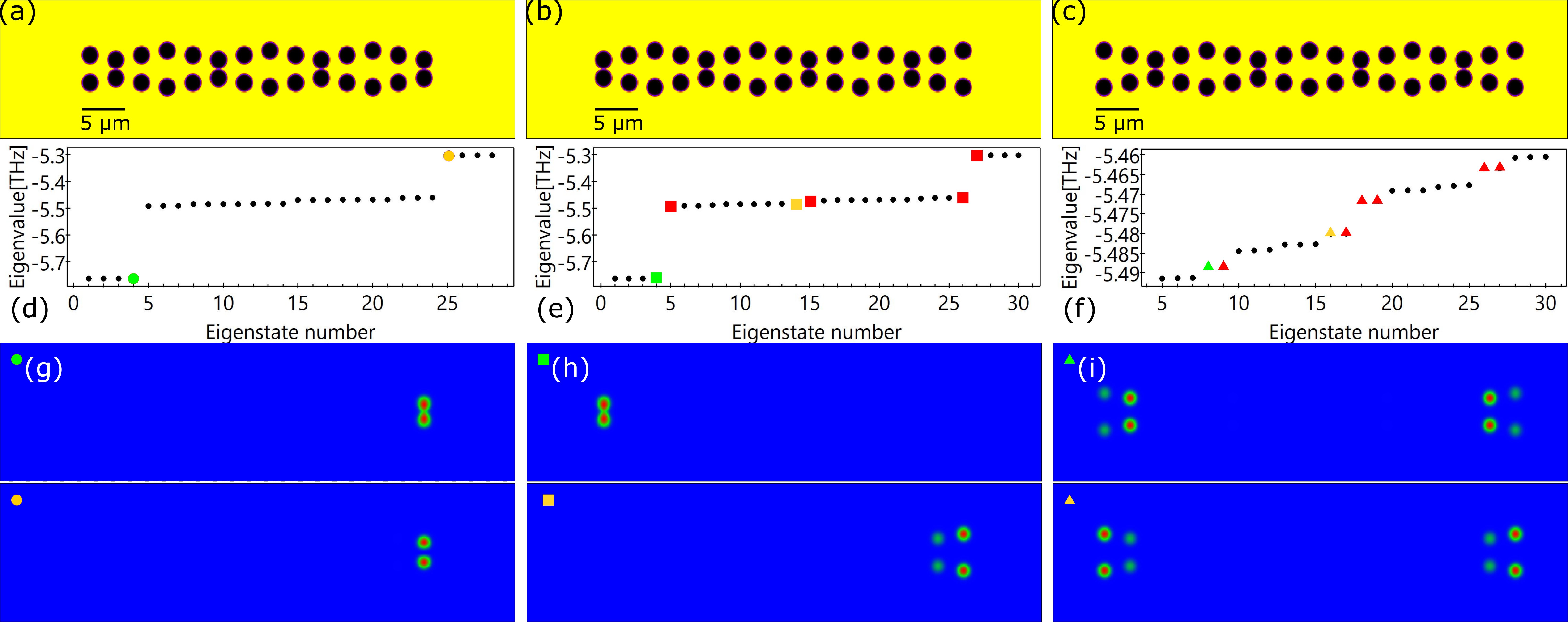}
  \caption{\textbf{Potential chains with different lengths and the corresponding edge states.} (a-c) Profiles of the double-wave chains with the same lattice parameters, $a=3$ $\mu$m, $A=2$ $\mu$m, and $d=0.5$ $\mu$m, but different lengths along the horizontal ($x$) direction. (d-f) Eigenfrequencies of the linear eigenstates versus the eigenstate numbers. (g-i) Selected edge states of the chains (a-c), corresponding to the symbols in (d-f), respectively. 
  }
  \label{fig:s3}
\end{figure}

To realize different edge states, one can also cut or add some potential wells in the chain. Here we consider the chain in Fig. 1(a) as the standard one. By cutting four potential wells at the right edge as shown in Fig. \ref{fig:s3}(a), similar edge states like the one shown in Fig. \ref{fig:s2}(h) form [see Fig. \ref{fig:s3}(d,g)]. If two potential wells at the left edge is removed [Fig. \ref{fig:s3}(b)], two types of edge states appear in the same chain with one at the left edge, where the potential wells are the least separated, and the other one at the right edge, where the potential wells are the most separated [Fig. \ref{fig:s3}(e,h)]. With adding two more potential wells to the left side of the chain to make it symmetric along $x$ direction [Fig. \ref{fig:s3}(c)], both edges are allowed to be occupied by the condensates with the same profile as shown in [Fig. \ref{fig:s3}(f,i)].

\newpage

\section*{III. Eigenstates at the inter-wave separation $A=2.5$ $\mu$m}
\begin{figure}[H]
  \centering
   \includegraphics[width=1\columnwidth]{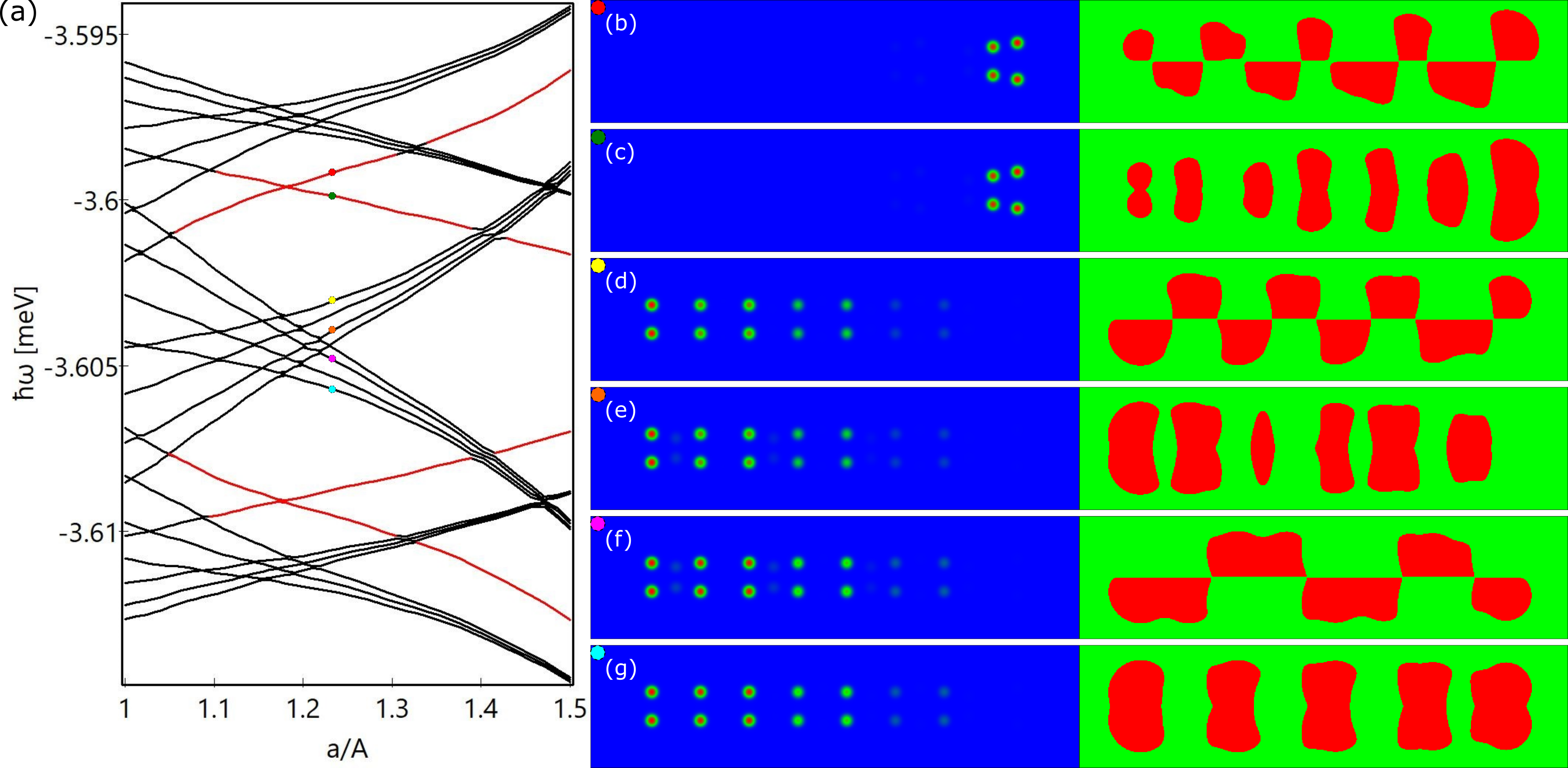}
  \caption{\textbf{Linear eigenstates.} (a) Dependence of the eigenenergies of the linear eigenstates in the double-wave chain on the inter-wave separation $A$, calculated based on the chain in Fig. 1(a) with $a=3$ $\mu$m, $d=0.5$ $\mu$m, and $\theta=0$. Red lines indicate the edge states and black lines are the bulk states. (b-g) Density (left) and phase (right) distributions of the selected eigenstates, corresponding to the symbols in (a), respectively.
  }
  \label{fig:s4}
\end{figure}

When the inter-wave separation $A=2.5$ $\mu$m, the density of the condensate of the edge states in the four potential wells at the right edge are almost identical [Fig. \ref{fig:s4}(b,c)]. Figure \ref{fig:s4}(d-g) shows the eigenstates (bulk states) with the energies below the two edge states. The condensates in such bulk states can be suppressed and consequently the condensate at the left edge is enhanced, forming new edge states, i.e., nonlinearity-induced edge states (see Fig. 4 in the main text). 

\newpage

\section*{IV. Multistable edge states}
\begin{figure}[H]
  \centering
   \includegraphics[width=0.9\columnwidth]{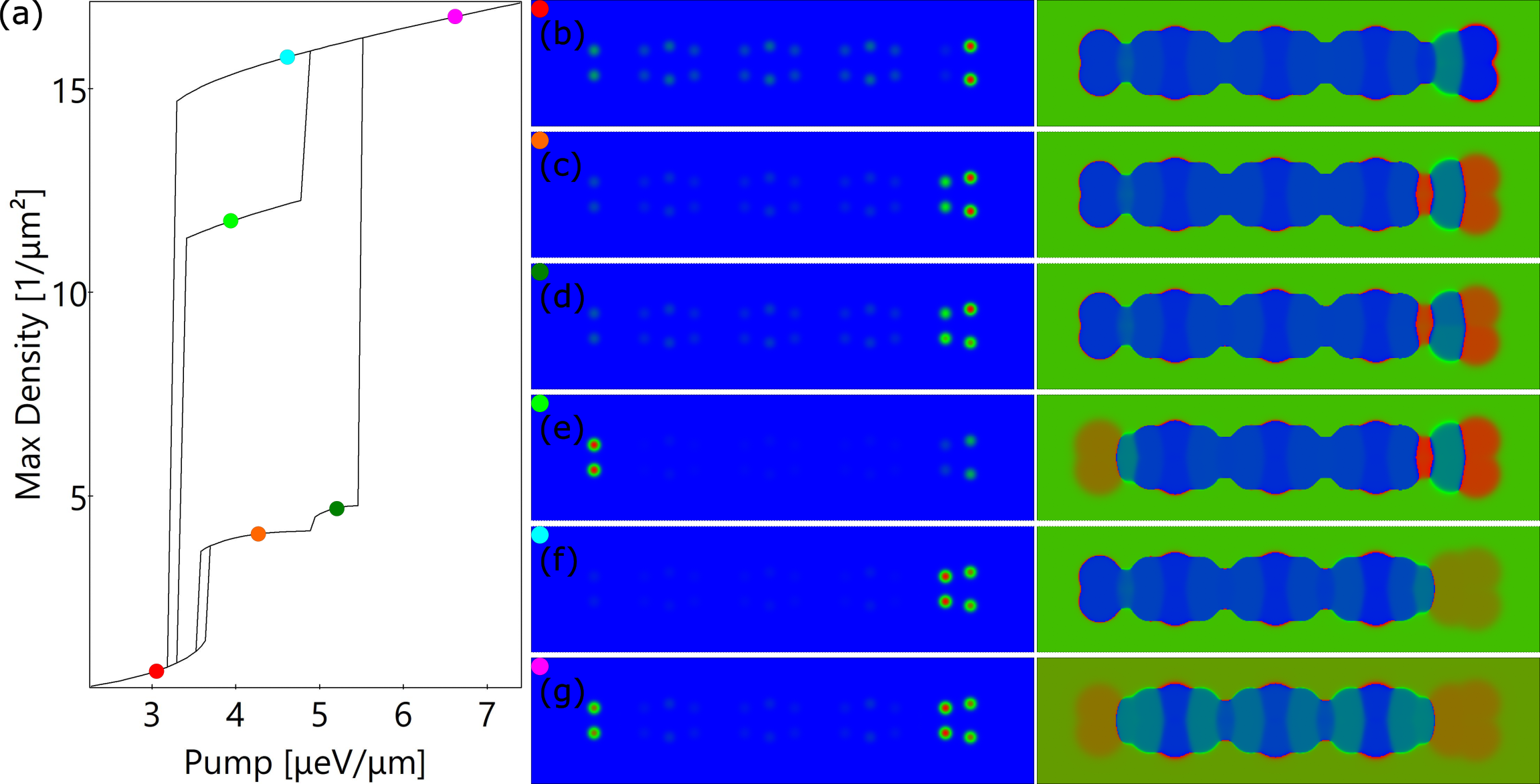}
  \caption{\textbf{Bistable edge states.} (a) Dependence of the peak density of the nonlinear edge states on the amplitudes of the pump with photon energy $\hbar\omega=$-3.5982 meV [the chain parameters are $a=3$ $\mu$m, $A=2$ $\mu$m, $d=0.5$ $\mu$m, and $\theta=0$, the same with the one shown in Fig. 1(a)]. (b-g) Density (left) and phase (right) profiles of the states marked in (a), respectively.
  }
  \label{fig:s5}
\end{figure}

From Fig. 1(b) or Fig. \ref{fig:s4}(a), one can see that when $a/A=1.5$, the edge states are more isolated from the bulk states, so that in this case, the edge state [Fig. \ref{fig:s5}(c,f)] can be solely excited without the excitation of the left edge state [c.f. Fig. 4(b)]. The left edge can still be excited when the pump intensity is stronger as shown in Fig. \ref{fig:s5}(a,e). When the pump intensity is stronger, the 0-$\pi$ edge state at the right edge is still slightly influenced by the $\pi$-$\pi$ state [see the density distribution in Fig. \ref{fig:s5}(d)], even there is a significant energy gap between them. It is clear that under some specific pump intensity range, the bistability remains in this case as shown in Fig. \ref{fig:s5}(a).

\newpage

\section*{V. Corner states}
\begin{figure}[H]
  \centering
   \includegraphics[width=1.0\columnwidth]{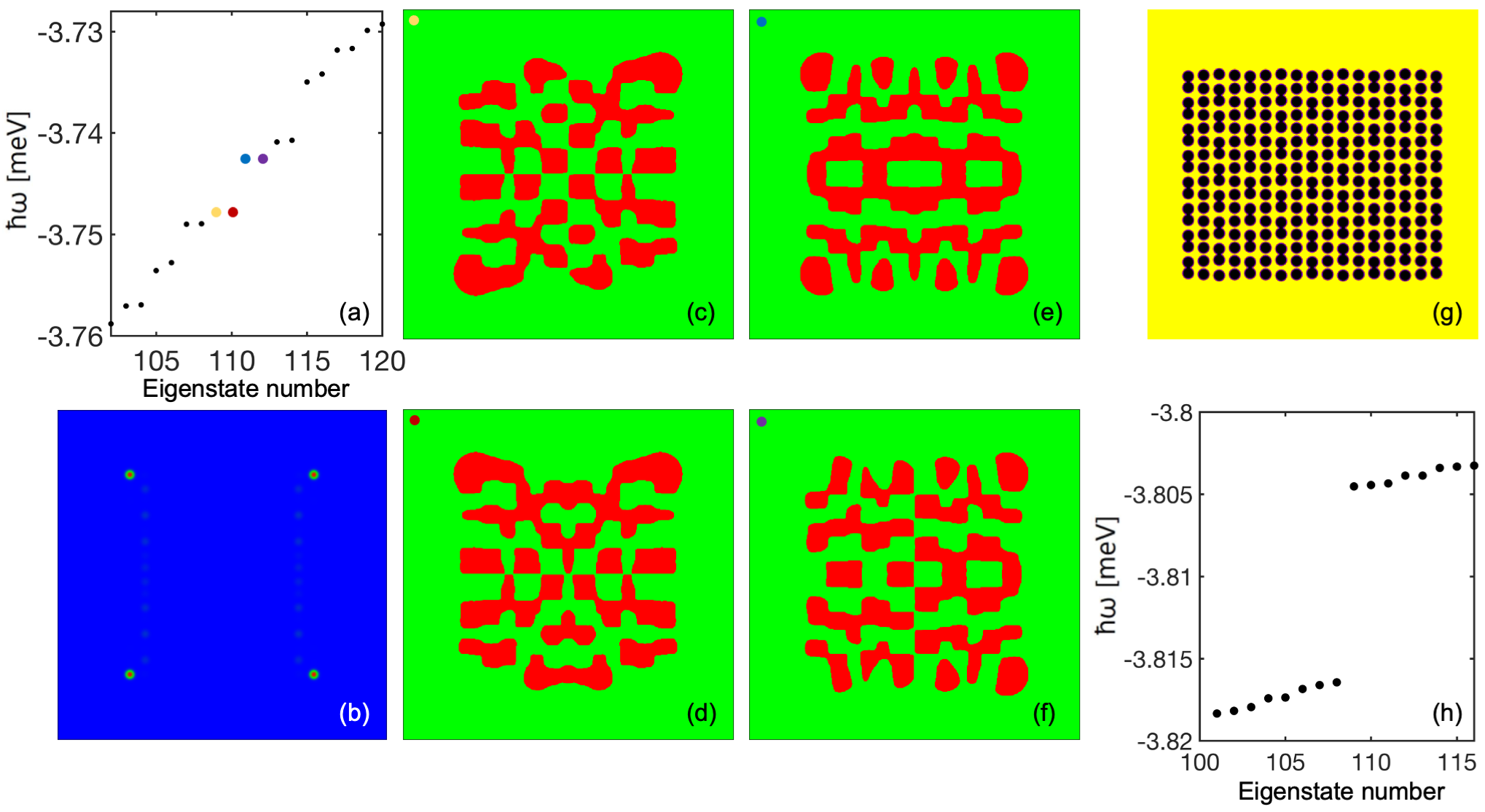}
  \caption{\textbf{Corner states.} (a) Selected linear eigenstates (energy vs. number) of the lattice in Fig. 5(b). (b) Density distribution of the corner state 109 in (a). (c-f) Phase profiles of the four corner states highlighted in (a). (g) A 2D multi-wave lattice pattern with $a=2.8$ $\mu$m, $A=2$ $\mu$m, and $d=0.2$ $\mu$m. (h) Selected linear eigenstates (energy vs. number) of the lattice in (g).
  }
  \label{fig:s6}
\end{figure}

It can be seen from Fig. \ref{fig:s6}(a) that the four corner states share two different energies. This is because there is a very small number of condensate being loaded into the chains next to the left and right edges [Fig. \ref{fig:s6}(b)]. The coupling of the condensates in the edge potentials and the potentials next to them induces the energy difference. The lower energy represents that the condensates in the adjacent vertical chains have the same phase [see Fig. \ref{fig:s6}(c,d)], while the higher energy indicates that their phase difference is $\pi$, see Fig. \ref{fig:s6}(e,f).

If a 2D lattice has the structure shown in Fig. \ref{fig:s6}(g), in the two SSH chains at the left and right edges, the intra-cell coupling is apparently stronger than the inter-cell coupling. Consequently, the left and right edges prevent the formation of the edge states along the vertical direction, so that there is no corner states in the corresponding band gap as can be seen in Fig. \ref{fig:s6}(h). 

\newpage

\begin{figure}[H]
  \centering
   \includegraphics[width=1.0\columnwidth]{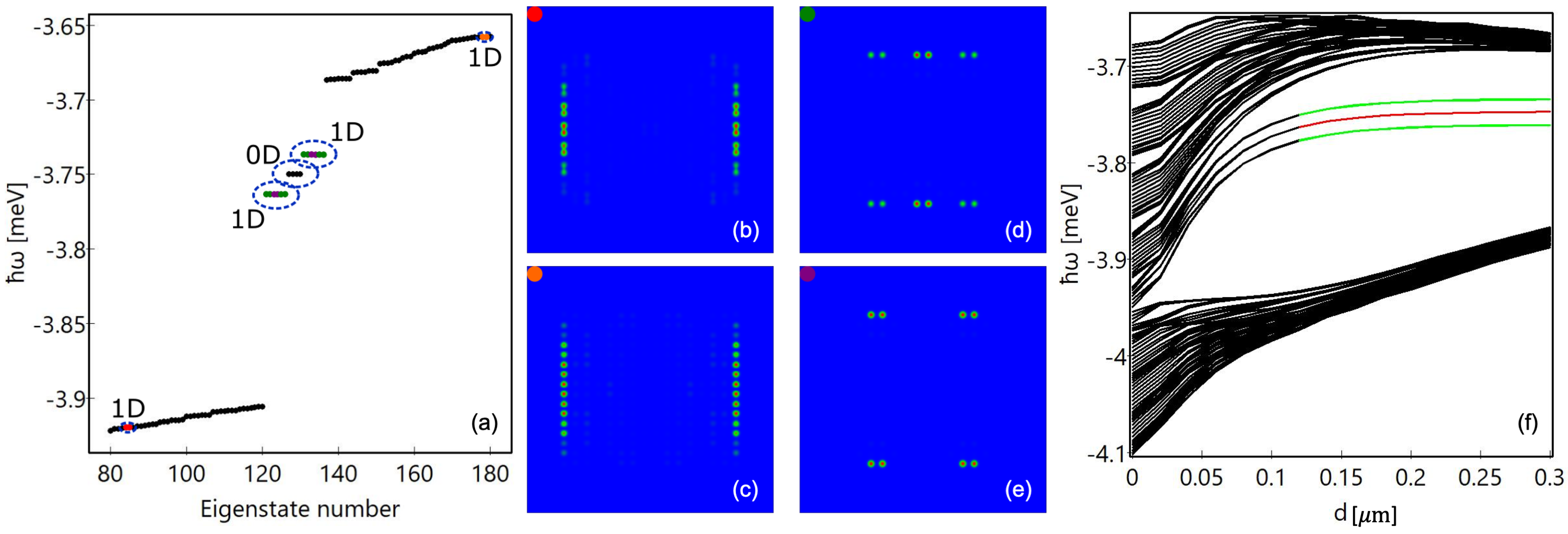}
  \caption{\textbf{Edge states in the 2D lattice.} (a) Selected linear eigenstates (energy vs. number) of the 2D lattice in Fig. 5(f). Here, 0D corner states and 1D edge states are marked. (b-e) Density distributuions of the 1D edge states highlighted in (a). (f) Dependence of the eigenenergies on the modulation amplitude $d$ in the 2D lattice with $A=2 \mu$m and $a=2.8 \mu$m fixed. Red lines indicate the 0D corner states, green lines indicate the 1D edge states, and black lines indicate the bulk states.}
  \label{fig:s6_2}
\end{figure}

In the 2D lattice presented in Fig. 5(f), there exist also 1D edge states as illustrated in Fig. \ref{fig:s6_2}. The 1D edge states at left and right edges [Fig. \ref{fig:s6_2}(b,c)] reside in the energy bands, inseparable from the bulk states, while the edge states appeared at upper and lower edges [Fig. \ref{fig:s6_2}(d,e)] are gapped. From the modulation amplitude $d$ dependent spectrum [Fig. \ref{fig:s6_2}(f)] one can see the transition of the edge and corner states from topologically trivial to nontrivial as $d$ increases.
\newpage

\section*{VI. Robustness of edge and corner states}
\begin{figure}[H]
  \centering
   \includegraphics[width=1\columnwidth]{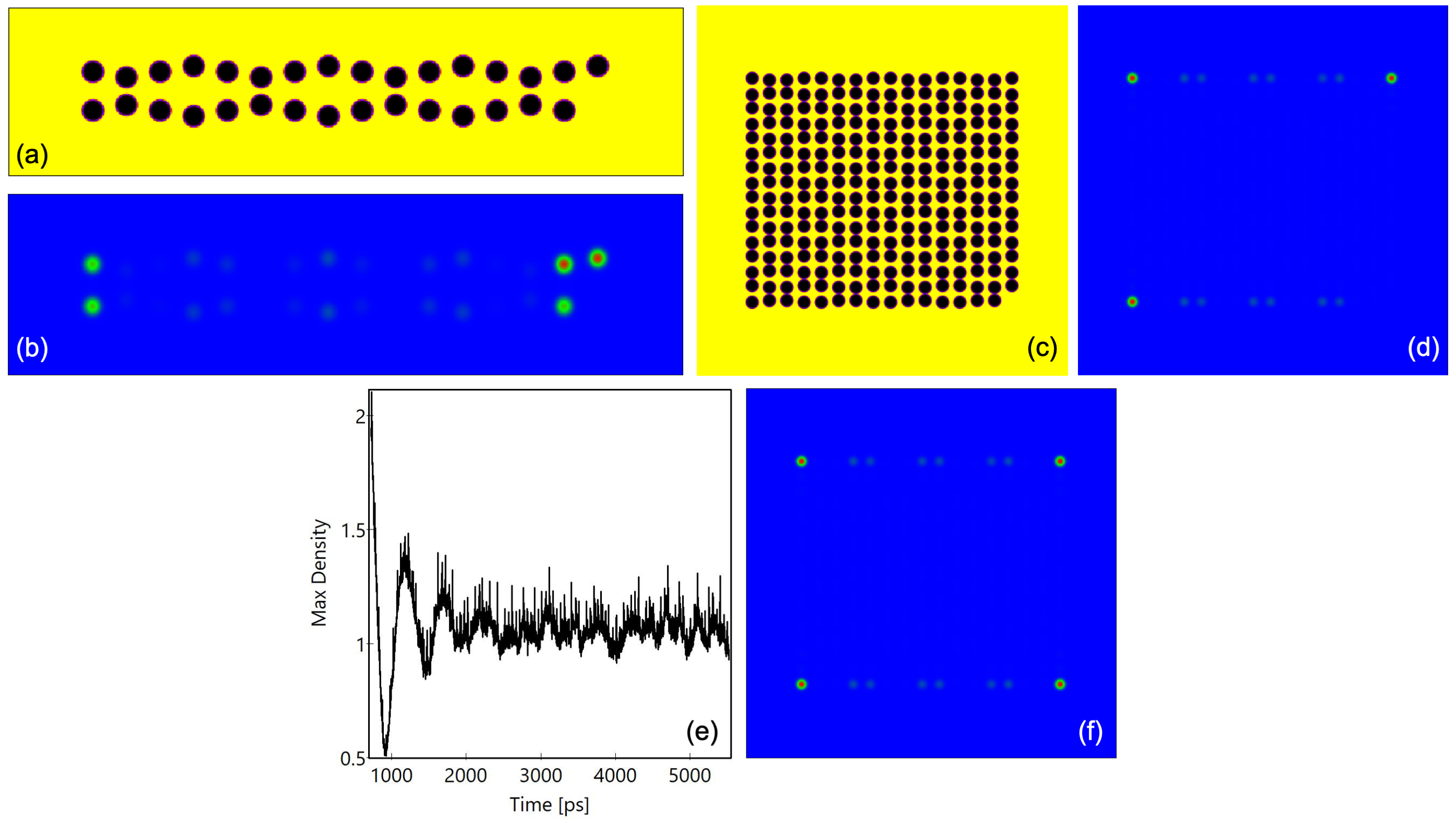}
  \caption{\textbf{Edge and corner states with potential defects.} (a) Profile of a DW chain with the same parameters of the one shown in Fig. 1(a) but a missing potential well at the right edge of the lower chain. (b) The corresponding density profile created by the same pump used in Fig. 4. (c) Profile of a 2D structure with the same parameters of the one shown in Fig. 5(f) but a missing potential well at the right bottom corner. (d) The corresponding density profile created by the same pump used in Fig. 6. (e) Time evolution of the peak density of the corner state shown in Fig. 6(b) with random white noise being added during the time evolution. (f) Density profile of the corner state at $t=5000$~ps.}
  \label{fig:s7}
\end{figure}

To demonstrate the robustness of the topological edge and corner states in the nonlinear regime, potential defects are applied to the original lattices. Figure \ref{fig:s7}(a) shows the 1D DW chain studied in the main text, but a potential well is removed from the lower wave chain. When we use the same excitation manner introduced in Fig. 4, one can see from Fig. \ref{fig:s7}(b) that the edge state, corresponding to the one shown in Fig. 4(b), can still be excited and is stable. This holds also for the 2D case as can be seen from Fig. \ref{fig:s7}(c,d) which correspond to the excitation studied in Fig. 6. These results are stable states obtained from the time evolution and evidence that the topological states studied in our work are very robust against strong defects. These topological states are also robust against significant white noise. An example is shown in Fig.~\ref{fig:s7}(e,f), starting from an initial condition that despite the addition of white noise leads to stable time evolution into the state in Fig.~\ref{fig:s7}(f).

\newpage

\section*{VII. Chern index of the even and odd subspaces}
In this section, the Chern index relation between the even and odd subspaces is derived. Following the formula used in the main text, the vertical couplings are $t+\lambda\cos(2\pi p /(N+1)j+\theta)$  and the horizontal couplings are $1$  . We know that the total Hamiltonian of two uncoupled AAH models reads:
$\hat{H}_{BD}=\begin{pmatrix}h_+ & 0 \\ 0 & h_-\end{pmatrix}$.
It is beneficial to write the position informed momentum space Hamiltonian of the odd subspace:
\begin{equation}
 h_-(k_x)=t+\begin{bmatrix} \Lambda_1 &e^{ik_x/N} & 0 & \dots &0&  e^{-ik_x/N} \\ e^{-ik_x/N} & \Lambda_2  \\ 0 & & & & \ddots\\\vdots &  & &\ddots & \ddots & e^{ik_x/N} \\ 0\\ e^{ik_x/N} & 0& \dots & 0 & e^{-ik_x/N} & \Lambda_{N}\end{bmatrix}\,.   
\end{equation}
In this equation, we have $\Lambda_j=\lambda\cos(2\pi p/(N+1)j+\theta)$. Furthermore for the even subspace we have, $h_+: h_- |\Lambda_j \rightarrow -\Lambda_j, t \rightarrow -t$, meaning the modulations are $\pi$ in phase between two subspaces. Next, eigenstates of the odd subspace satisfy $h_- |\Psi_{-,j}\rangle=E_{-,j}|\Psi_{-,j}\rangle$. It is apparent we have the following equivalence relation: $h_-(k_x)|\Psi_{-,j}(k_x)\rangle=-h_+(k_x+N\pi)|\Psi_{-,j}(k_x)\rangle=E_{-,j}(k_x)|\Psi_{-,j}(k_x)\rangle $, which leads us to equivalence: 
\begin{equation}
\begin{aligned}
    E_{+,j}(k_x)=-E_{-,N-j+1}(k_x+N\pi)\,, \\
    |\Psi_{+,j}(k_x)\rangle=|\Psi_{-,N-j+1}(k_x+N\pi)\rangle\,.
\end{aligned}
\end{equation}
This relation allows us to conclude $C_{+,j}=C_{-,N-j+1}$.


The Chern index can be calculated separately for the $h_+$ and $h_-$ sectors since they are decoupled by mirror symmetry. The Chern index for the $j$-th band is expressed as
\begin{equation}
  C_{\pm ,j}=\frac{1}{2\pi}\int_{-\pi}^{\pi}dk_x\int_{0}^{2\pi}d\theta \,\text{Im}(\langle \partial_{k_x}\Psi_{\pm,j}|\partial_{\theta}\Psi_{\pm,j}\rangle-\langle \partial_{\theta}\Psi_{\pm,j}|\partial_{k_x}\Psi_{\pm,j}\rangle).
\end{equation}
In practice, we numerically evaluated the Chern index for each mirror-symmetric sector, in the tight-binding approximation, according to the well established numerical method
[T. Fukui \textit{et al.}, Journal of the Physical Society of Japan, 74, 1674-1677 (2005).]. The result is consistent with the appearance of edge states, hence the bulk-edge correpondance [Y. Hatsugai \textit{et al.}, Physical Review Letters, 71, 3697 (1993).]. The Chern index is the preferred topological index, since in the 2D parameter space $(k_x,\theta)$, the time-reversal symmetry links $(k_x,\theta)$ to $(-k_x,-\theta)$. Despite the actual physical structure is time-reversal invariant under fixed $\theta$. This allows us to define the 2D Chern index for such AAH-like models. Previous studies have shown that the AAH model allows the observation of topological pumping, which is determined by the Chern index [Y. Ke \textit{et al.}, Laser \& Photonics Reviews, 10, 995-1001 (2016).].

\newpage

\section*{VIII. Bulk-edge correspondence}
\begin{figure}[H]
  \centering
   \includegraphics[width=0.55\columnwidth]{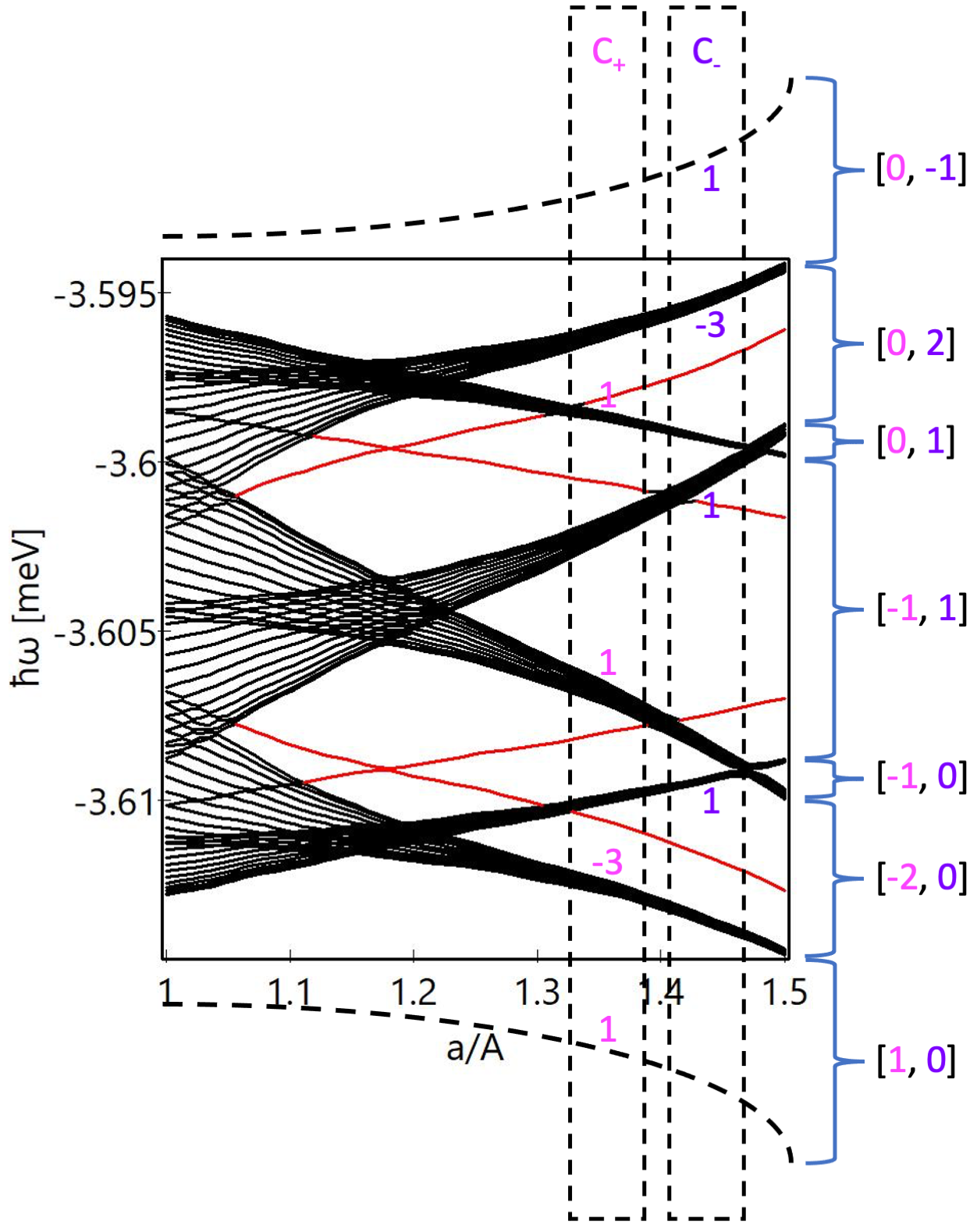}
  \caption{\textbf{Bulk-edge correspondence and the corresponding Chern numbers of the band gaps.} The energy bands are the same with that in Fig. 1(b), while the two dashed lines below and above represent the other two energy bands at much smaller and larger energies, respectively.}
  \label{fig:s8}
\end{figure}

The Chern numbers of each band gap can be calculated through the bulk-edge correspondence as illustrated in Fig. \ref{fig:s8}. For each energy band, there is a Chern number which has been calculated and given in the main text, and they have been labeled to the corresponding energy bands in Fig. \ref{fig:s8}. The Chern numbers of a band gap can be calculated by accumulating the corresponding numbers of all the energy bands below it. For example, the top gap's Chern numbers [0, -1] is obtained by applying $C_{+}=1+1-3+1$ and $C_{-}=-3+1+1$. Note that the edge states, denoted by solid red lines, seems absent in the top gap, despite the nontrivial Chern index. This is due to the fact that these edge states appear at $\theta \approx 0.75\pi$ (Fig.\ref{fig:s2}), while Fig. \ref{fig:s8} is calculated with $\theta =0$. 
